\documentclass[10pt,twocolumn,letterpaper]{article}

\usepackage{cvpr}              

\usepackage[dvipsnames]{xcolor}

\definecolor{cvprblue}{rgb}{0.21,0.49,0.74}
\usepackage[pagebackref,breaklinks,colorlinks,citecolor=cvprblue]{hyperref}
\usepackage{graphicx}
\usepackage{amsmath}
\usepackage{amssymb}
\usepackage{booktabs}
\usepackage{algorithm,algorithmic}
\DeclareMathOperator*{\minimize}{minimize}
\usepackage{multirow}
\usepackage{float}

\title{Universal Perturbation-based Secret Key-Controlled Data Hiding}

\author{Donghua Wang\\
Zhejiang University\\
{\tt\small wangdonghua@zju.edu.cn}
\and
Wen Yao\thanks{Corresponding author}\\
Chinese Academy of Military Science\\
\and
Tingsong Jiang$^*$\\
Chinese Academy of Military Science
\and
Xiaoqian Chen\\
Chinese Academy of Military Science
}

\begin{document}
\maketitle
\begin{abstract}
Deep neural networks (DNNs) are demonstrated to be vulnerable to universal perturbation, a single quasi-perceptible perturbation that can deceive the DNN on most images. However, the previous works are focused on using universal perturbation to perform adversarial attacks, while the potential usability of universal perturbation as data carriers in data hiding is less explored, especially for the key-controlled data hiding method. In this paper, we propose a novel universal perturbation-based secret key-controlled data-hiding method, realizing data hiding with a single universal perturbation and data decoding with the secret key-controlled decoder. Specifically, we optimize a single universal perturbation, which serves as a data carrier that can hide multiple secret images and be added to most cover images. Then, we devise a secret key-controlled decoder to extract different secret images from the single container image constructed by the universal perturbation by using different secret keys. Moreover, a suppress loss function is proposed to prevent the secret image from leakage. Furthermore, we adopt a robust module to boost the decoder's capability against corruption. Finally, A co-joint optimization strategy is proposed to find the optimal universal perturbation and decoder. Extensive experiments are conducted on different datasets to demonstrate the effectiveness of the proposed method. Additionally, the physical test performed on platforms (e.g., WeChat and Twitter) verifies the usability of the proposed method in practice. 
\end{abstract}    
\section{Introduction}
\label{sec:intro}

\begin{figure}[t]
	\centering
	\begin{minipage}{1.\linewidth}
		\centering
		\includegraphics[width =1.\linewidth]{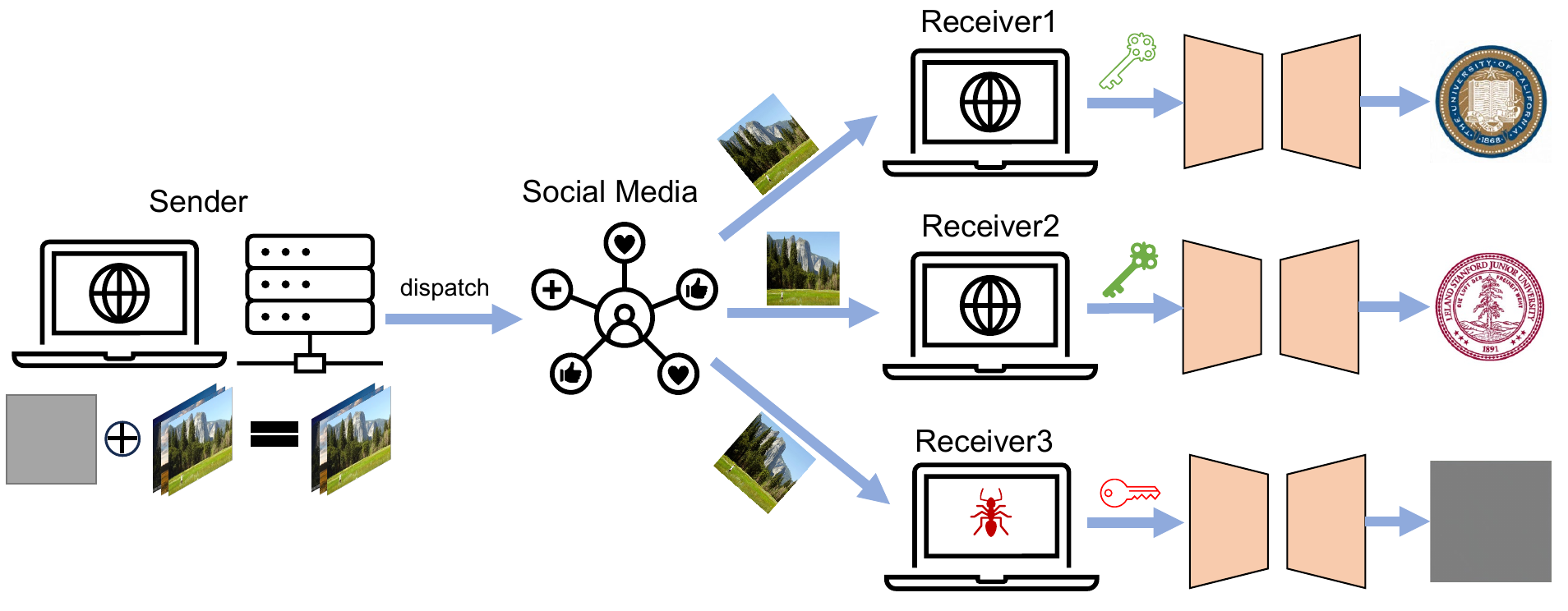}
	\end{minipage}
\caption{Multi-message dispatch scenario based on the proposed method. The sender constructs the container image by adding a universal perturbation to any cover image and then dispatches it to the website. Different receivers can decode their desired secret image using their secret keys. }
\label{fig:first_page}
\end{figure}

Although deep neural networks (DNNs) achieve notable progress in various computer vision tasks such as image classification \cite{he2016deep,dosovitskiy2021an}, object detection \cite{ren2015faster,redmon2016you}, and semantic segmentation \cite{long2015fully,he2017mask}, DNNs have shown their fragile to adversarial examples \cite{li2022approximated,wang2022fca,li2023adaptive,wang2023rfla}, which are crafted by a small malicious perturbation that is invisible to human beings but can mislead the DNN to output wrong results. Recently, Zhang \etal \cite{zhang2020understanding} claimed that the universal adversarial perturbation (UAP) \textit{contains features independent of the images to attack}, which means that UAP possesses the class-related information that determines the model's decision. Inspired by that, we deem that universal perturbation can not only perform adversarial attacks but also have potential positive usage, which is less explored.

Regarding the positive usage of adversarial perturbation, some researchers have made attempts, such as adversarial training \cite{shafahi2019adversarial,wong2020fast} and auxiliary model decisions \cite{salman2021unadversarial,wang2022defensive}. Adversarial training is an effective regularization method \cite{goodfellow2015fgsm} to enhance the model's adversarial robustness by introducing the adversarial perturbation to the training set \cite{shafahi2019adversarial}. In the auxiliary model decision, universal perturbation is treated as the defensive patch \cite{salman2021unadversarial,wang2022defensive} to enhance the model's capability of recognizing the image that may suffer corruption caused by bad weather (e.g., thick fog and snow). However, the adversarial training method treats the perturbation as a special noise distribution, while the auxiliary method engenders conspicuous adversarial patterns. By contrast, in this work, we explore the potential usage of universal perturbation as the secret carrier in data hiding.


Data hiding has long been an important research field and attract a wide range of attention. Especially, in the recently year, data hiding enjoys the tremendous development of deep learning, emerging many impressive methods \cite{zhu2018hidden,zhang2020understanding,tancik2020stegastamp,lu2021large,jing2021hinet,xu2022robust}. In general, the DNN-based data hiding method trains the data hiding network and secret decode network to realize the message encoding and extraction, respectively. Although universal perturbation had been used to implement data hiding \cite{zhang2020understanding}, they were required to train a hiding network to generate a universal perturbation, which is regarded as a secret data carrier. Moreover, their method can only decode single secret data from a single container, limiting their usage in some fields, e.g., message dispatch, in which different receivers can decode different messages using different secret keys. 



In this paper, we propose a novel universal perturbation based secret-key controlled data hiding method. Specifically, we encode multiple secret images to a single universal perturbation, which can be added to arbitrary cover images to produce the container image, and the decoded secret image depends on the secret key used during decoding. Unlike the previous work that devised a hiding network to encode the secret image, we adopt the Adam optimizer to find the single universal perturbation directly to hide the multi-secret image. Then, we devise a secret-key controlled decoder network to extract the secret image, where the decoder takes the container image and secret key as input and outputs the corresponding secret image. To protect the secret image from being decoded by using the illegal secret key, we devise a suppress loss term to guide the decoder to map the illegal secret key to the wrong secret image. Additionally, we exploit a robust module to improve the robustness of the encoder. Extensive quantitative and quantity results verify the effectiveness of the proposed method. Finally, we provide a concrete usage scenario of the proposed method in multi-message dispatch in Figure \ref{fig:first_page}, a sender constructs the container image with only a single universal perturbation, then dispatches it on the website, and then the different receivers can decode different secret images by using different secret keys.

Our contribution is summarized as follows.

\begin{itemize}
\item We propose a novel universal perturbation based secret key-controlled data hiding method, implementing a single universal perturbation to hide multiple secret images, and different secret images can be decoded from the same container image by using different secret keys.
\item We devise a suppress loss function to prevent the leakage of the secret image and leverage a robustness module to enhance the robustness of the decoder.
\item We conduct the physical experiment by dispatching the container image over the social media platform and verifying the potential usage of our method in real applications.
\end{itemize}
\section{Related Works}
In this section, we briefly introduce the universal adversarial perturbation and the DNN-based data hiding. 

\subsection{Universal Adversarial Perturbation}
Universal adversarial perturbation (UAP) is a single perturbation that can deceive the DNN model on most images. The existing UAP approaches can be grouped into optimized-based and generative model-based methods. The former cover the most approach, such as Moosavi \etal \cite{moosavi2017universal} proposed the first UAP method, which iterative accumulates the perturbation for every training image. Mopuri \etal \cite{mopuri2017fast,mopuri2018generalizable} devised a novel loss function that updates the UAP to maximize the activation value of the network and lead to wrong output. Liu \etal \cite{liu2019universal} leveraged the uncertainty to optimize UAP. Zhang \etal \cite{zhang2020understanding} explored the mutual information among UAP and clean images and is used to update the UAP. Follow-up, Zhang \etal \cite{zhang2021data} devised a cosine similarity loss to optimize UAP by minimizing the logit out between the clean image and UAP perturbed images. Recently, Li \etal \cite{li2022learning} proposed to learn the UAP from the image-specific perturbation under the guidance of consistency regularation. By contrast, generative-based methods cite{mopuri2018nag,mopuri2018ask} usually adopt the generative model to generate a distribution of UAP, such as Mopuri \etal \cite{mopuri2018ask} proposed a data-free UAP method, which extracted the class impression image as the training image, and trained a Generative Adversarial Network (GAN) to generate UAP. The methods mentioned above use the universal perturbation to perform the adversarial attack, while the positive usage of universal perturbation is less explored.

\subsection{DNN-based Data Hiding}
The existing DNN-based data hiding approaches include two-fold: steganography and watermarking. Both of them aim to hide specific information in other types of media; the former focuses on the concealment of the encoded message, the accuracy of the decoded message, and avoiding detection, while the latter concentrates on the identification of the creator that may be corrupted. For steganography, Baluja \etal \cite{baluja2017hiding,baluja2019hiding} first utilized a hiding network to hide the data (i.e., RGB image) to another image, and then a reveal network was utilized to extract the image. Zhang \etal \cite{zhang2019steganogan} utilized a GAN to hide the binary message into images, whereas a critic network is devised to generate a more realistic encoded image. Jing \etal \cite{jing2021hinet} devised an invertible network to simultaneously realize the image hiding and extracting. Similar to \cite{zhang2019steganogan}, Wei \etal \cite{wei2022generative} utilized the GAN-based method to hide the binary data in the image, where the difference is that the critic network is replaced by a steganalyzer. For watermarking, Tancik \etal \cite{tancik2020stegastamp} developed a method to hide the URL link in physical photographs. Jia \etal \cite{jia2021mbrs} adopted the differentiable JEPG compression to realize a robustness watermarking technique. In addition, some research attempts to develop a method to implement steganography and watermarking simultaneously. Zhu \etal \cite{zhu2018hidden} first devised an end-to-end trainable network for hiding binary data, and an adversary network was devised to improve the encoded image quality. Furthermore, they introduced a set of robustness techniques to realize the robustness watermark. Zhang \etal \cite{zhang2020udh} proposed a universal deep data hiding method that embeds the one image data to different images. However, the existing DNN-based data-hiding method requires a hiding network to hide the single secret in the cover image, while our method instead uses a universal perturbation to hide multi-secret in the cover image. Moreover, the decoded secret image can be controlled by the decoding secret key. 
\section{Methodology}

\begin{figure*}[ht]
	\centering
	\begin{minipage}{1.\linewidth}
		\centering
		\includegraphics[width =1.\linewidth]{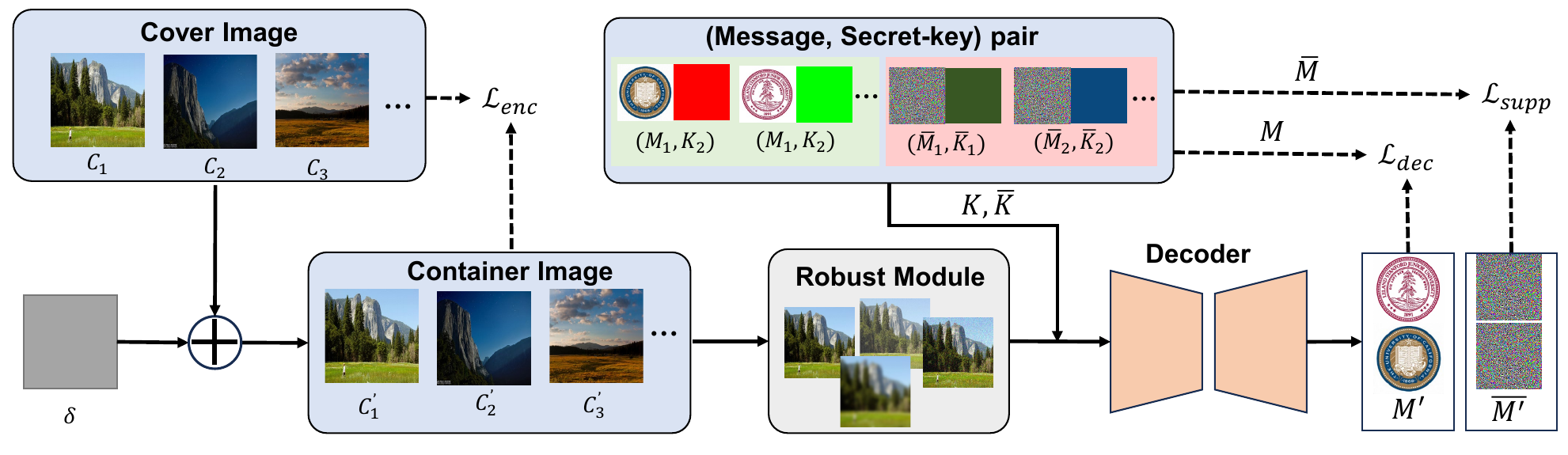}
	\end{minipage}
\caption{Overview of the proposed secret key-controlled universal perturbation data hiding framework. The framework consists of two steps. In the first step, one universal perturbation $\delta$ is added to any cover image $C$, producing container image $C'$; $\mathcal{L}_{enc}$ is devised to optimize $\delta$ by minimizing the discrepancy between $C$ and $C'$, leading to invisible perturbation. In the second step, the container image is first disrupted by the used robust module and then fed into the decoder together with secret key image $K$, engender the decoded secret image $M'$. $\mathcal{L}_{dec}$ is designed to extract correct secret image. Moreover, $\mathcal{L}_{supp}$ is devised to avoid decoding the secret image with the illegal secret key.}
\label{fig:framework}
\end{figure*}

\subsection{Problem Statement}
In this work, we treat the universal perturbation as the secret data carrier for data hiding and controlling the decoded secret image by the decoding secret key. Specifically, given cover images $C$ and universal perturbation $\delta$ that hides secret images $M$, the container image $C'$ can be obtained by
\begin{equation}
C' = C + \delta.
\end{equation}
And a decoder $\mathcal{D}_\theta$ with weights of $\theta$ is devised to extract secret images $M'$ from container image $C'$ by using corresponding decoding secret keys $K$, i.e., 
\begin{equation}
\label{eq:decode}
M'_i = \mathcal{D}_\theta(C', K_i), i \in 1,2,...,N,
\end{equation}
where $N$ is the number of secret images hidden by universal perturbation $\delta$.

The optimization object is engendering the universal perturbation $\delta$ for most cover images $C$ with the cost of smaller modification on cover images and making the decoder correctly extract the secret images with the provided decoding secret key from the container image, which can be expressed as 

\begin{equation}
\min_{\delta, \theta} ||M-M'||, ~~ s.t., ~ ||\delta||_p \leq \epsilon, for ~most ~~ C,
\end{equation}
where $||\cdot||$ is the distance metric, $||\delta||_p$ is the $L_p$-norm used to bound the $\delta$ be in the $\epsilon$ radius of the sphere for imperceptible, and $\delta$ can be used for most cover images $C$. 

The framework of the proposed method is illustrated in Figure \ref{fig:framework}, which is composed of universal perturbation and the secret-key controlled decoder. We will elaborate as follows.

\subsection{Universal Perturbation as Secret Data Carrier}
Universal adversarial perturbation is a single quasi-imperceptible perturbation \cite{moosavi2017universal,mopuri2018generalizable}, which was used to perform adversarial attacks against the DNN previously. However, Zhang \etal \cite{zhang2020understanding} stated that the universal perturbation carries class-related information in the DNN-based classifier, which motivates us to explore whether the universal perturbation can serve as a secret data carrier for data hiding. Specifically, multiple secret images are encoded in the universal perturbation, which can be used to add to most cover images to construct container images. Unlike the universal adversarial perturbation used to attack the classification model to modify the classification result, the universal perturbation constructed container image can be correctly decoded by a decoder which is a regression model. Therefore, we optimize the universal perturbation $\delta$ by solving the following constrained optimization problem

\begin{equation}
\min \delta,~~ s.t.~~\mathcal{D}_\theta(C + \delta, K) = M,~  ||\delta||_p \leq \epsilon,
\end{equation}
where $L_p$ is set to $L_\infty$-norm, which implemented by $\max(\min(\delta, -\epsilon), \epsilon))$. 

\subsection{Secret-key Controlled Decoder}
The decoder in previous work \cite{zhu2018hidden,zhang2020udh} only extracts a single secret image from a container image. In contrast, we devised a secret-key-controlled decoder in this work to extract the different secret images from the same container image with different decoding secret keys. Specifically, we devise a secret-key-controlled decoder $\mathcal{D}_\theta$ consisting of two components: a secret-key decoder $\mathcal{D}_{sk}$ and a main decoder $\mathcal{D}$. The former is devised to project the secret key to the embedding space, which has the same resolution as the container image. Then, the embedded secret key is channel-wise concatenated with the container image and fed into the main decoder. Furthermore, we introduce a differentiable robust module $\mathcal{R}$ to enhance the decoder's capability against various corruption (e.g., JPEG compression, blur). Therefore, the decoder can be optimized by

\begin{equation}
\min_\theta || \mathcal{D}_\theta(\mathcal{R}(C+\delta), \mathcal{D}_{sk}(K_i)) - M_i||,~i \in 1,2,...,N,
\end{equation}
where the $M_i \in M$ and $K_i \in K$, which comes in pair, i.e., using $K_i$ to extract the $M_i$, $N$ denotes that the number of $(K_i,M_i)$ pair. 

\subsection{Optimization}
To realize the goals mentioned above, we devised the loss function and co-joint training strategy, which will be exhaustively introduced as follows. 

\textbf{Loss Function.} The first loss function is devised to generate the universal perturbation with a smaller magnitude (namely $\mathcal{L}_{enc}$) to avoid introducing conspicuous traces on the cover image during the construction of the container image. Therefore, we adopt the following MSE-based loss function to achieve this goal.

\begin{equation}
\label{eq:l_enc}
\mathcal{L}_{enc} = MSE(C+\delta, C).
\end{equation}

The second decoder loss function ($\mathcal{L}_{dec}$) is designed to train the decoder to extract the correct secret image from the container image that may be corrupted. Similarly, we adopt the following MSE-based loss function to minimize the discrepancy between the decoded secret image and the real one.

\begin{equation}
\label{eq:l_dec}
\mathcal{L}_{dec} = MSE(M'_i, M_i).
\end{equation}

Moreover, given the high dimensionality of the embedding space, it is inevitable to decode the secret image with the illegal secret key. To protect the secret image from leakage, we devise a suppress loss function $\mathcal{L}_{supp}$ to guide the decoder to map illegal secret keys to the nonsense secret image (can also be replaced with misleading secret images). Specifically, we randomly sample the illegal secret key $\overline{K}$ and nonsense secret image $\overline{M}$ pair and minimize the discrepancy between the illegal secret image and the corresponding decoded secret image at each iteration. We adopt the MSE as the $\mathcal{L}_{supp}$ as well, which is expressed as

\begin{equation}
\label{eq:l_supp}
\mathcal{L}_{supp} = MSE(\mathcal{D}(\mathcal{R}(C+\delta), \mathcal{D}_{sk}(\overline{K}_i)), \overline{M}_i).
\end{equation}

\textbf{Co-joint training strategy}
Unlike the universal adversarial attack, which has a target pre-trained model, we have no pre-trained model. To this end, we have to co-joint train the universal perturbation and decoder. Furthermore, given the two loss mentioned above terms are relevant minimization problems, we solve them by co-joint training. Finally, the optimization object is expressed as

\begin{equation}
\minimize_{\delta, \theta}~~ \mathcal{L}_{enc} + \mathcal{L}_{dec} + \lambda \mathcal{L}_{supp},
\end{equation}
where $\lambda$ is the weight to balance the decoder training. Note that we set the same weight for $\mathcal{L}_{enc}$ and $\mathcal{L}_{dec} + \lambda \mathcal{L}_{supp}$ as they are two distinct tasks. The Algorithm \ref{alg1:algorithm1} describes the training procedure in detail.

\begin{algorithm}[t]
	\caption{Training procedure of the proposed secret key-controlled universal perturbation data hiding method}
	\label{alg1:algorithm1}
	\textbf{Input}: cover image $C$, the decoder $\mathcal{D}$ with weights $\theta$, $(M, K)$ pair, batch size $bs$, max iteration $MaxIter$\\
	\textbf{Output}: Optimal $\delta^*, \theta^*$
	\begin{algorithmic}[1] 
		\STATE Initialize the $\delta$ and $\theta$
		\FOR {$itr=i,...,MaxIter$}
			\STATE Random sample $bs$ cover images $\left\{C_1,...,C_{bs}\right\}$ from $C$, legal secret image and key pair $\left\{(M_1,K_1),...,(M_{bs},K_{bs})\right\}$ from $(M,K)$
			\STATE Random spawn illegal secret image and key pair $\left\{(\overline{M}_1,\overline{K}_1),...,(\overline{M}_{bs},\overline{K}_{bs})\right\}$.
			\STATE Add the universal perturbation $\delta$ to cover images and obtain $\left\{C_1+\delta, ..., C_{bs}+\delta\right\}$ 
    		\STATE Use the decoder $\mathcal{D}_\theta$ to obtain the decoded image $\left\{M'_1, ...,M'_{bs}\right\}$ and $\left\{\overline{M}'_1,...,\overline{M}'_{bs}\right\}$ by Equation \ref{eq:decode} 
			\STATE Updating the $\delta$ with Equation \ref{eq:l_enc}
			\STATE Updating the $\theta$ with Equation \ref{eq:l_dec} and Equation \ref{eq:l_supp}
		\ENDFOR
	\end{algorithmic}
\end{algorithm}

\begin{table*}[ht]
\centering
\setlength\tabcolsep{2pt}
\caption{Quantitative result of universal perturbation under different $\epsilon$ on different datasets. }
\label{tab:main_result}
\begin{tabular}{|c|c|ccccccc|cccccc|}
\hline
\multirow{2}{*}{}                           & \multirow{2}{*}{$\epsilon$} & \multirow{2}{*}{Dataset} & \multicolumn{6}{c|}{Key-Stanford}                                       & \multicolumn{6}{c|}{Key-UCLA}                         \\ \cline{4-15} 
                                            &                      &  & SSIM$\uparrow$            & PSNR$\uparrow$           & FID$\downarrow$            & LPIPS$\downarrow$              & APD$\downarrow$    & MSE$\downarrow$ & SSIM$\uparrow$            & PSNR$\uparrow$           & FID$\downarrow$            & LPIPS$\downarrow$              & APD$\downarrow$   & MSE$\downarrow$      \\ \hline
\multicolumn{1}{|c|}{\multirow{8}{*}{C-C$'$}} & \multirow{2}{*}{6}   & S2W          & 0.9544 & 39.83 & 7.55  & 0.003615 & 2.11 & 0.000104 & 0.9544 & 39.83 & 7.55  & 0.003615 & 2.11 & 0.000104 \\
\multicolumn{1}{|c|}{}                      &                      & ImageNet       & 0.9549 & 39.81 & 6.65  & 0.00138  & 2.12 & 0.000105 & 0.9549 & 39.81 & 6.65  & 0.00138  & 2.12 & 0.000105 \\ \cline{2-15} 
\multicolumn{1}{|c|}{}                      & \multirow{2}{*}{8}  & S2W             & 0.9261 & 37.37 & 11.42 & 0.005504 & 2.81 & 0.000183 & 0.9261 & 37.37 & 11.42 & 0.005504 & 2.81 & 0.000183 \\
\multicolumn{1}{|c|}{}                      &                      & ImageNet       & 0.9266 & 37.36 & 9.96  & 0.002014 & 2.82 & 0.000184 & 0.9266 & 37.36 & 9.96  & 0.002014 & 2.82 & 0.000184 \\ \cline{2-15}
\multicolumn{1}{|c|}{}                      & \multirow{2}{*}{10}  & S2W            & 0.8949 & 35.49 & 15.25 & 0.007365 & 3.5  & 0.000282 & 0.8949 & 35.49 & 15.25 & 0.007365 & 3.5  & 0.000282 \\
\multicolumn{1}{|c|}{}                      &                      & ImageNet       & 0.8955 & 35.48 & 13.1  & 0.002614 & 3.51 & 0.000283 & 0.8955 & 35.48 & 13.1  & 0.002614 & 3.51 & 0.000283 \\ \cline{2-15}
\multicolumn{1}{|c|}{} 						& \multirow{2}{*}{12}   & S2W           & 0.8623 & 33.96 & 18.95 & 0.009156 & 4.19 & 0.000402 & 0.8623 & 33.96 & 18.95 & 0.009156 & 4.19 & 0.000402 \\
\multicolumn{1}{|c|}{}                      &                      & ImageNet       & 0.8629 & 33.94 & 16.04 & 0.003182 & 4.2  & 0.000404 & 0.8629 & 33.94 & 16.04 & 0.003182 & 4.2  & 0.000404 \\ \midrule\hline\noalign{\smallskip} 
\multicolumn{1}{|c|}{\multirow{8}{*}{M-M$'$}}& \multirow{2}{*}{6} & S2W             & 0.9892 & 32.82 & 31.74 & 0.001313 & 3.23 & 0.000566 & 0.9865 & 34.81 & 10.15 & 0.001246 & 2.46 & 0.000371 \\
\multicolumn{1}{|c|}{}                      &                      & ImageNet       & 0.9885 & 32.94 & 32.08 & 0.000464 & 3.21 & 0.000594 & 0.9855 & 34.91 & 10.58 & 0.000426 & 2.49 & 0.000433 \\ \cline{2-15}
\multicolumn{1}{|c|}{}                      & \multirow{2}{*}{8}  & S2W             & 0.995  & 37.57 & 11.04 & 0.000529 & 2.05 & 0.000196 & 0.9948 & 40    & 5     & 0.000426 & 1.58 & 0.00014  \\
\multicolumn{1}{|c|}{}                      &                      & ImageNet       & 0.9958 & 38.31 & 5.78  & 0.000086 & 1.75 & 0.00018  & 0.9915 & 37.37 & 10.28 & 0.000283 & 1.96 & 0.000223 \\ \cline{2-15}
\multicolumn{1}{|c|}{}                      & \multirow{2}{*}{10}  & S2W            & 0.9928 & 35.51 & 24.33 & 0.000937 & 2.42 & 0.00029  & 0.9908 & 37.02 & 8.68  & 0.0011   & 2.09 & 0.000232 \\
\multicolumn{1}{|c|}{}                      &                      & ImageNet       & 0.9954 & 38.53 & 11.91 & 0.000164 & 1.95 & 0.000179 & 0.9941 & 39.44 & 5.54  & 0.000176 & 1.68 & 0.000165 \\ \cline{2-15}
\multicolumn{1}{|c|}{}                      & \multirow{2}{*}{12}  & S2W            & 0.9967 & 39.85 & 6.16  & 0.00024  & 1.43 & 0.000123 & 0.9949 & 39.27 & 6.45  & 0.000472 & 1.37 & 0.000145 \\
\multicolumn{1}{|c|}{}                      &                      & ImageNet       & 0.9959 & 38.5  & 7.34  & 0.000093 & 1.7  & 0.000157 & 0.9944 & 40.02 & 5.25  & 0.000152 & 1.65 & 0.000125 \\ \cline{1-15}
\end{tabular}
\end{table*}
\section{Experiments}
In this section, we first briefly introduce the experiment settings and then perform extensive experiments to verify the effectiveness of the proposed method.
\subsection{Settings}
\textbf{Datasets.} We evaluate the proposed method on two datasets: ImageNet \cite{deng2009imagenet} and summer2winter (abbr. S2W) released by \cite{zhu2017unpaired}. For ImageNet, we randomly sample 10000 and 1000 images from the ImageNET training and validation set as our training and test set, respectively. For summer2winter, the training set includes 1231 images, and the test set includes 309 images. In our experiment, the resolution of the image is set to 400 $\times$ 400.

\textbf{Evaluation Metrics.} To comprehensive evaluate the proposed method, we adopt six evaluation metrics from three aspects, containing two traditional subjective measurement metrics: SSIM and PSNR, these two metrics the larger, the better ($\uparrow$); and two novel network-based metrics: FID \cite{heusel2017gans} and LPIPS \cite{zhang2018unreasonable}, these two metrics the smaller, the better ($\downarrow$); and two quantitative metrics APD proposed by \cite{zhang2020udh} and MSE, these two metrics, the smaller the better($\downarrow$). We calculate the above metrics using the implementation provided in \cite{pytorch2020ignite} except LPIPIS\footnote{https://github.com/S-aiueo32/lpips-pytorch.}, APD, and MSE. 

\textbf{Comparison Methods.} As we are the first ones to use the universal perturbation as the secret 
adat carrier and devise a secret-key controlled decoder, similar work is less. UDH \cite{zhang2020udh} is the only method similar to ours, but they used a hiding network to generate the universal perturbation for data hiding and were unable to decode different secret images from the same container image.

\textbf{Implementation Details.} Given that the message is embedded into the universal perturbation, we only have to devise the secret-key-controlled decoder. The decoder contains two components pre-decoder and main decoder. The pre-decoder is constructed by five convolution layers, where the middle three layers are the residual blocks. The main decoder is a UNet-architecture with skip connect, consisting of five convolutional and upsample layers. We adopt the Adam optimizer to optimize the universal perturbation and the decoder with a learning rate 1e-4. The batch size is set to 64, and the max iteration is 140K. Unless specific, we follow \cite{moosavi2017universal,zhang2021data} set the $\epsilon$ to $10/255$, $\lambda$ is set to 0.05. The number of secret images is set to two, i.e., the UCLA and Stanford school badge, and the corresponding secret keys are the pure color images for simplicity, i.e., red for UCLA and green for Stanford. All codes are implemented in PyTorch \footnote{Code will be released after being published.}, and all experiments are conducted on an NVIDIA RTX3090ti GPU (24GB) cluster.

\subsection{Quantitative result}
\begin{figure*}[t]
	\centering
	\begin{minipage}{.8\linewidth}
		\centering
		\includegraphics[width =1.\linewidth]{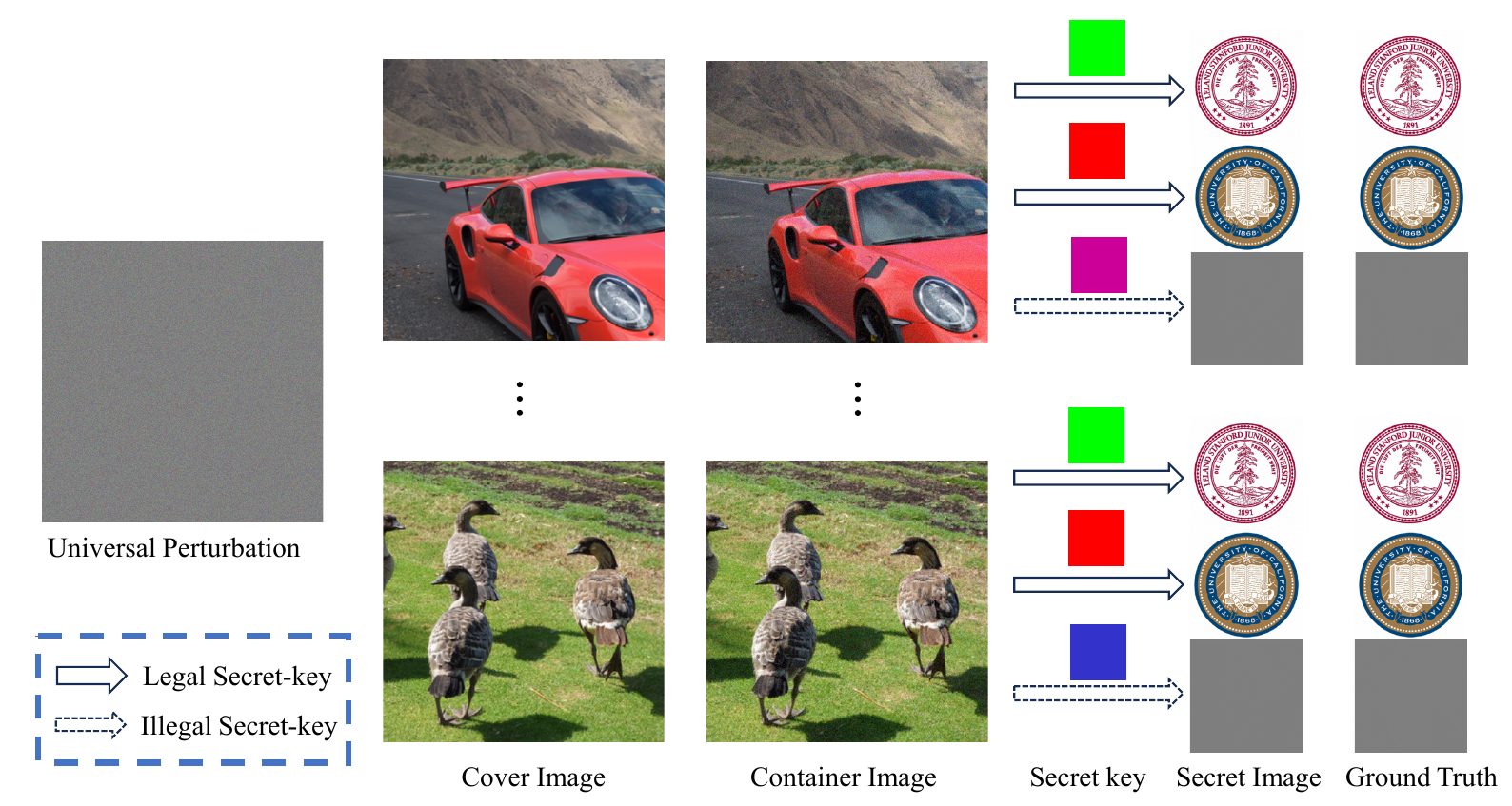}
	\end{minipage}
\caption{Visualization examples.}
\label{fig:examples}
\end{figure*}

\begin{table*}[h]
\centering
\setlength\tabcolsep{2pt}
\caption{Comparison results of Ours and UDH \cite{zhang2020udh} in terms of various metrics on different datasets.}
\label{tab:comparison_result}
\begin{tabular}{clccccccccccccc}
\hline 
\multirow{2}{*}{Dataset}       & \multirow{2}{*}{Method} & \multicolumn{6}{c}{C-C$'$}      &           & \multicolumn{6}{c}{M-M$'$}      \\ \cline{3-8} \cline{10-15} 
                               &                         & SSIM$\uparrow$            & PSNR$\uparrow$           & FID$\downarrow$            & LPIPS$\downarrow$              & APD$\downarrow$            & MSE$\downarrow$                &           & SSIM$\uparrow$            & PSNR$\uparrow$           & FID$\downarrow$            & LPIPS$\downarrow$              & APD$\downarrow$            & MSE$\downarrow$               \\ \hline
\multirow{2}{*}{S2W} & UDH \cite{zhang2020udh}   & 0.8599          & 28.15          & 43.86          & 0.0135           & 7.85          & 0.00179   &       & 0.9831          & 32.11          & 17.03          & 0.00141          & 3.76          & 0.00065          \\
                               & OUR     & \textbf{0.8949} & \textbf{35.49} & \textbf{15.25} & \textbf{0.00737} & \textbf{3.5}  & \textbf{0.00028} & & \textbf{0.9918} & \textbf{36.27} & \textbf{16.51} & \textbf{0.00102} & \textbf{2.26} & \textbf{0.00026} \\ \hline
\multirow{2}{*}{ImageNet}  & UDH \cite{zhang2020udh}  & \textbf{0.9216} & 33.1           & 17.93          & 0.00261          & 4.25          & 0.00059      &    & 0.9918          & 36.08          & \textbf{5.33}  & 0.00018          & 2.48          & 0.00032  \\
                               & OUR    & 0.8955          & \textbf{35.48} & \textbf{13.1}  & 0.00261          & \textbf{3.51} & \textbf{0.00028} & & \textbf{0.9963} & \textbf{38.99} & 8.73  & \textbf{0.00017} & \textbf{1.82} & \textbf{0.00017} \\ \hline
\end{tabular}
\end{table*}

In this section, we investigate the performance of the proposed method under different maximum allowable modifications. Specifically, we conduct the experiment on the S2W and ImageNet datasets with the $\epsilon$ in four magnitudes, i.e., $\epsilon \in \left\{6,8,10,12\right\}/255$. Then, the secret image is extracted from the same container dataset by using two different secret keys (e.g., Key-Stanford and Key-UCLA) and obtaining two secret image sets. Finally, the evaluation metric is calculated. 

The evaluation results are listed in Table \ref{tab:main_result}. As we can observe, on the one hand, the image quality of two different secret images extracted by the decoder using two secret keys is high. For example, at the smallest modification magnitude of $\epsilon$=6, the minimum PSNR between the decoded secret image and ground truth (abbrev. M-M$'$) on two datasets exceeds 33 dB (i.e., 33.82 dB on S2W), which suggests that the proposed method can balance the concealment of container images and extraction accurate of secret images. Correspondingly, the container image's image quality has a lower bound that depends on the universal perturbation's maximum magnitude. On the other hand, we observed a trend with the change of $\epsilon$: with the increase of $\epsilon$, the evaluation metrics of C-C$'$ are increasingly worse, while the M-M$'$ is getting better. Such a phenomenon is expected, as the increasing of $\epsilon$, the more conspicuous trace appears in universal perturbation, which can be easily learned by the decoder and produce better results. Moreover, we also observe the inconsistency between LPIPS and other evaluation metrics (e.g., $\epsilon$=12), such as the evaluation metrics of M-M$'$ on S2W are better than ImageNet while worse in LPIPS, demonstrating the necessary of multi measurement. 

In addition, we provide the visualization example randomly sampled from the ImageNet in Figure \ref{fig:examples}. As we can see, the single universal can be added to arbitrary images to construct the container image, which is visually similar to the cover image. Then, with the trained decoder, different secret images can be extracted from the same container image using different secret key images. Moreover, using the illegal secret key image unable to extract the correct secret image, which demonstrates the effectiveness of the devised suppress loss function.

\subsection{Comparison result}
To perform the comparison, we reproduce the UDH \cite{zhang2020udh} on our dataset and use the hiding network in UDH to generate two universal perturbations for Key-UCLA and Key-Stanford, respectively. Then calculate the evaluation metric on two secret key constructed datasets, respectively. For simplicity, we average the evaluation results of two secret keys. Table \ref{tab:comparison_result} reports the comparison results. As we can observe, on the one hand, the image quality of the decoded secret image of the proposed method outperforms the UDH significantly on all evaluation metrics on both datasets. On the other hand, we are slightly falling behind in PSNR, FID, and LPIPS of C-C$'$ than UDH, which can be attributed to the hiding network used in UDH engendering a small magnitude of the universal perturbation. However, their cost is the poor image quality of decoded secret images. In contrast, the proposed method realizes the secret hiding by substituting the hiding network with a single universal perturbation with an upper bound magnitude. Despite the slightly worse image quality of container images, we obtain better image quality of decoded secret images. Moreover, the proposed method can significantly save the store cost and increase usability in practice.

\subsection{Robustness evaluation}
In this section, we investigate the robustness of the proposed method against different corruptions. Specifically, we perform the Gaussian blur with different kernel sizes (i.e., 3, 5, 7, 9) and JPEG compresses with different quality factors (i.e., 10, 30, 50, 70, 90) on the container image, respectively, and then decoded them by the decoder with the legal secret key. We average the evaluation result of two secret keys for simplicity. Table \ref{tab:robust_gaussian} and Table \ref{tab:robust_jpeg} list the evaluation results on Gaussian blur and JPEG compression. As we can observe, under the corruption of the Gaussian blur, the image quality of the extracted secret image drops significantly compared to the original one, and the largest decline occurs in the kernel size of 5, where the drop magnitude of PSNR is 9.57dB on the S2W dataset. As for JPEG compression, the image quality of the decoded secret image is acceptable except in the case that the quality factor is 10 on the S2W dataset. The PSNR under different quality factors approaches 30 dB. As discussed, the proposed method is more robust to JPEG compression than the Gaussian blur. Despite the decline in the evaluation metric, the content of the secret image is recognizable, and please refer to Appendix \ref{supp:robust}.

\begin{table}[]
\centering
\setlength\tabcolsep{2pt}
\caption{Robustness evaluation on Gaussian blur with different kernel size (ks). "-" indicates the original results.}
\label{tab:robust_gaussian}
\begin{tabular}{c|c|c|ccccc}
\hline
                       & ks             &          & SSIM$\uparrow$ & PSNR$\uparrow$   & FID$\downarrow$& LPIPS$\downarrow$  & APD$\downarrow$     \\ \hline
\multirow{2}{*}{C-C$'$}  & \multirow{2}{*}{-}    & S2W      & 0.8949 & 35.49 & 15.25 & 0.007365 & 3.51   \\ 
                       &                         & ImageNet & 0.8955 & 35.48 & 13.1  & 0.002614 & 3.51   \\ \hline
\multirow{10}{*}{M-M$'$} & \multirow{2}{*}{-}    & S2W      & 0.9918 & 36.26 & 16.51 & 0.001019 & 2.25   \\
                       &                         & ImageNet & 0.9947 & 38.98 & 8.73  & 0.00017  & 1.81   \\ \cline{2-8}
                       & \multirow{2}{*}{3}      & S2W      & 0.9262 & 27.72 & 51.02 & 0.006012 & 6.36   \\
                       &                         & ImageNet & 0.9557 & 30.33 & 44.27 & 0.00124  & 4.71   \\ \cline{2-8}
                       & \multirow{2}{*}{5}      & S2W      & 0.8912 & 26.69 & 68.37 & 0.008175 & 8.84   \\
                       &                         & ImageNet & 0.83   & 26.33 & 97.07 & 0.004038 & 12.22  \\ \cline{2-8}
                       & \multirow{2}{*}{7}      & S2W      & 0.9079 & 27.26 & 61.15 & 0.007058 & 7.65   \\
                       &                         & ImageNet & 0.8453 & 26.62 & 87.23 & 0.003724 & 11.34  \\ \cline{2-8}
                       & \multirow{2}{*}{9}      & S2W      & 0.9137 & 27.46 & 58.78 & 0.006677 & 7.2    \\
                       &                         & ImageNet & 0.8506 & 26.73 & 85.59 & 0.003643 & 10.97  \\ \hline
\end{tabular}     
\end{table}
\begin{table}[]
\centering
\setlength\tabcolsep{2pt}
\caption{Robustness evaluation on JPEG compression with different quality factors(Q). "-" indicates the original results.}
\label{tab:robust_jpeg}
\begin{tabular}{c|c|c|ccccc}
\hline
                         & Q               &              & SSIM$\uparrow$ & PSNR$\uparrow$   & FID$\downarrow$& LPIPS$\downarrow$  & APD$\downarrow$     \\ \hline
\multirow{2}{*}{C-C$'$}  & \multirow{2}{*}{-}  & S2W      & 0.8949 & 35.49 & 15.25 & 0.007365 & 3.5   \\
                         &                     & ImageNet & 0.8955 & 35.48 & 13.1  & 0.002614 & 3.51  \\ \hline
\multirow{12}{*}{M-M$'$} & \multirow{2}{*}{-}  & S2W      & 0.9918 & 36.26 & 16.51 & 0.001019 & 2.25  \\
                       &                       & ImageNet & 0.9947 & 38.98 & 8.73  & 0.00017  & 1.81  \\ \cline{2-8}
                       & \multirow{2}{*}{10}   & S2W      & 0.8532 & 20.54 & 70.57 & 0.010782 & 10.98 \\ 
                       &                       & ImageNet & 0.9482 & 29.48 & 35.88 & 0.001325 & 5.58  \\ \cline{2-8}
                       & \multirow{2}{*}{30}   & S2W      & 0.9686 & 28.1  & 23.59 & 0.002359 & 4.18  \\
                       &                       & ImageNet & 0.9807 & 33.94 & 17.02 & 0.000507 & 3.27  \\ \cline{2-8}
                       & \multirow{2}{*}{50}   & S2W      & 0.9772 & 29.64 & 19.9  & 0.001733 & 3.57  \\
                       &                       & ImageNet & 0.9894 & 35.22 & 11.99 & 0.000288 & 2.63  \\ \cline{2-8}
                       & \multirow{2}{*}{70}   & S2W      & 0.9801 & 30.42 & 18.73 & 0.001522 & 3.34  \\
                       &                       & ImageNet & 0.9912 & 35.56 & 10.84 & 0.000244 & 2.48  \\ \cline{2-8}
                       & \multirow{2}{*}{90}   & S2W      & 0.9795 & 30.78 & 20.92 & 0.00165  & 3.39  \\
                       &                       & ImageNet & 0.9832 & 32.68 & 16.65 & 0.00045  & 3.19   \\ \hline
\end{tabular}
\end{table}

\subsection{Physical test}
\begin{figure}[t]
	\centering
	\begin{minipage}{.45\linewidth}
		\centering
		\includegraphics[width =1.\linewidth]{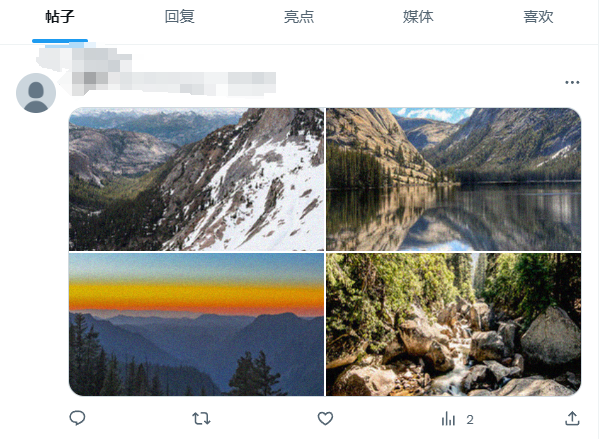}
		\centerline{\footnotesize Twitter}
	\end{minipage}
	\begin{minipage}{.45\linewidth}
		\centering
		\includegraphics[width =1.\linewidth]{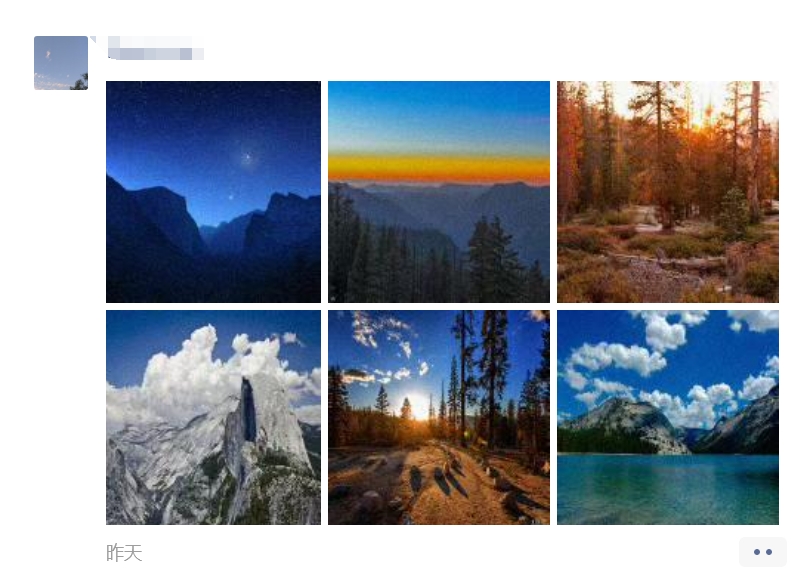}
		\centerline{\footnotesize WeChat}
	\end{minipage}
\caption{Container image in the social platform.}
\label{fig:physical}
\end{figure}

The usability of the data-hiding method is significant for their potential real-world application. In this section, we investigate the usability of the proposed method by exploring whether the decoder can extract the secret image from the container image that is transmitted to and saved from the social platform (i.e., WeChat and Twitter). Specifically, we randomly select 50 container images from the S2W dataset, then manually post them to the social platform (illustrated in Figure \ref{fig:physical}) and save them. Then, we average the evaluation results of two secret keys,  and Table \ref{tab:physical_test} reports the average result. As we can observe, the proposed method performs well on Twitter, and the evaluation metrics are slightly worse than the one decoded from no corruptions, where the drop magnitude is 0.5 dB (36.26 dB $\rightarrow$ 35.76 dB). In comparison, the semantic detail of the decoded result from the WeChat saved container image is significantly lost, and the PSNR drops 11.26 dB to 25dB, which is analogous to the JPEG compression with a quality factor of nearly 20. Despite the image detail is not satisfying, the content is recognizable as illustrated in Appendix \ref{supp:physical_test} for visualization examples.

\begin{table}[h]
\centering
\setlength\tabcolsep{2pt}
\caption{Physical evaluation results on WeChat and Twitter.}
\label{tab:physical_test}
\begin{tabular}{ccccccc}
\hline
        & SSIM$\uparrow$ & PSNR$\uparrow$   & FID$\downarrow$& LPIPS$\downarrow$  & APD$\downarrow$ & MSE$\downarrow$    \\ \hline
WeChat  & 0.9228 & 25    & 52.37 & 0.02544 & 6.68 & 0.00407 \\
Twitter & 0.9907 & 35.76 & 18.72 & 0.00503 & 2.34 & 0.00028 \\ \hline
\end{tabular}
\end{table}

\begin{table*}[ht]
\setlength\tabcolsep{2pt}
\caption{Effectiveness of the devise $\mathcal{L}_{supp}$.}
\label{tab:ablation_loss}
\begin{tabular}{clccccccccccccc}
\hline
\multirow{2}{*}{}     & \multirow{2}{*}{Dataset} & \multicolumn{6}{c}{without $\mathcal{L}_{supp}$}   &     & \multicolumn{6}{c}{with $\mathcal{L}_{supp}$}    \\ \cline{3-8} \cline{10-15} 
                      &     & SSIM$\uparrow$ & PSNR$\uparrow$   & FID$\downarrow$& LPIPS$\downarrow$  & APD$\downarrow$ & MSE$\downarrow$  &  & SSIM$\uparrow$   & PSNR$\uparrow$ & FID$\downarrow$   & LPIPS$\downarrow$  & APD$\downarrow$  & MSE$\downarrow$               \\ \hline
\multirow{2}{*}{C-C$'$} & S2W            & 0.8949 & 35.49 & 15.25  & 0.007365 & 3.5   & 0.000282 & & 0.8949 & 35.49 & 15.25  & 0.007365 & 3.5    & 0.000282 \\
                      & ImageNet                 & 0.8955 & 35.48 & 13.1   & 0.002614 & 3.51  & 0.000283 & &0.8955 & 35.48 & 13.1   & 0.002614 & 3.51   & 0.000283 \\ \hline
\multirow{2}{*}{M-M$'$} & S2W            & 0.6052 & 16.33 & 193.75 & 0.016471 & 35.79 & 0.068547 &  &0.3502 & 7.37  & 498.44 & 0.051024 & 102.62 & 0.183168 \\
                      & ImageNet                 & 0.5977 & 17.23 & 203.81 & 0.01165  & 36.34 & 0.069538 & & 0.3505 & 7.36  & 527.36 & 0.016494 & 102.7  & 0.183496 \\ \hline
\end{tabular}
\end{table*}

\begin{table}[H]
\caption{Influence of the number of secret images.}
\label{tab:ablation_num}
\setlength\tabcolsep{2pt}
\begin{tabular}{cc|cccccc}
\hline
Num                &      & SSIM$\uparrow$ & PSNR$\uparrow$   & FID$\downarrow$& LPIPS$\downarrow$  & APD$\downarrow$ & MSE$\downarrow$     \\ \hline
\multirow{2}{*}{2} & C-C$'$ & 0.8949 & 35.49 & 15.25 & 0.00737 & 3.5  & 0.00028 \\
                   & M-M$'$ & 0.9964 & 40.96 & 8.19  & 0.00043 & 1.45 & 0.00009 \\ \cline{1-8}
\multirow{2}{*}{3} & C-C$'$ & 0.8949 & 35.49 & 15.25 & 0.00737 & 3.51 & 0.00028 \\
                   & M-M$'$ & 0.9929 & 38.48 & 10.84 & 0.00059 & 1.79 & 0.00016 \\ \cline{1-8}
\multirow{2}{*}{4} & C-C$'$ & 0.8949 & 35.49 & 15.25 & 0.00737 & 3.5  & 0.00028  \\
                   & M-M$'$ & 0.99   & 35.5  & 13.87 & 0.00087 & 2.3  & 0.00034 \\ \cline{1-8}
\multirow{2}{*}{5} & C-C$'$ & 0.8949 & 35.49 & 15.25 & 0.00737 & 3.5  & 0.00028 \\ 
                   & M-M$'$ & 0.9827 & 33.01 & 33.51 & 0.00178 & 3.19 & 0.00058  \\ \hline
\end{tabular}
\end{table}

\begin{table}[H]
\setlength\tabcolsep{2pt}
\caption{Influence of secret image type.}
\small
\label{tab:ablation_image_type}
\begin{tabular}{cccccccc}
\hline
                        &        & SSIM$\uparrow$ & PSNR$\uparrow$   & FID$\downarrow$& LPIPS$\downarrow$  & APD$\downarrow$ & MSE$\downarrow$      \\ \hline
\multirow{2}{*}{Nature} & C-C$'$ & 0.8949 & 35.49 & 15.25 & 0.00737 & 3.5  & 0.00028 \\
                        & M-M$'$ & 0.8605 & 29.21 & 27.07 & 0.00968 & 7.14 & 0.00178 \\ \hline
\multirow{2}{*}{Text}   & C-C$'$ & 0.8949 & 35.49 & 15.25 & 0.00737 & 3.5  & 0.00028 \\
                        & M-M$'$ & 0.9984 & 40.6  & 0.74  & 0.0001  & 1.53 & 0.00009  \\ \hline
\multirow{2}{*}{Badge}  & C-C$'$ & 0.8949 & 35.49 & 15.25 & 0.00737 & 3.5  & 0.00028 \\
                        & M-M$'$ & 0.9964 & 40.96 & 8.19  & 0.00043 & 1.45 & 0.00009 \\ \hline
\end{tabular}
\end{table}

\subsection{Ablation study}
In this section, we conduct the ablation study to investigate the influence of important factors, including the effectiveness of the suppress loss, the secret image number, and the secret image type.

\textbf{Effectiveness of the $\mathcal{L}_{supp}$.}  In the preliminary experiment, we find that the decoder trained without using suppress loss can extract the correct secret image by using the illegal secret key, which is unacceptable in practice. Thus, we devise a loss function $\mathcal{L}_{supp}$ to solve this issue. To show the effectiveness of $\mathcal{L}_{supp}$, we rerun the experiment with/without using $\mathcal{L}_{supp}$. Then, as the decoded secret image extracted by the random secret key has no reference secret image, we calculate the evaluation metrics between the secret image extracted by the random secret key and the Key-UCLA and Key-Stranford, respectively. We average the results for simplicity, which are listed in Table \ref{tab:ablation_loss}. As we can observe, M-M$'$ is significantly influenced by $\mathcal{L}_{supp}$. For example, the PSNR is 17.23 dB on ImageNet without using $\mathcal{L}_{supp}$, while the PSNR is only 7.36 dB with using $\mathcal{L}_{supp}$, which means that the decoded secret image is unrecognized, achieving the goal of protecting secret images. We also provide the visualization example in Appendix \ref{supp:loss_supp}.

\textbf{Influence of the secret image number.} To explore how many secret images the universal perturbation can hide deserves further investigation, we investigate the performance of the proposed method on S2W when setting the number of secret images to $\left\{2,3,4,5\right\}$. We average the decoded results extracted by different secret keys, which are listed in Table \ref{tab:ablation_num}. As we can observe, on the one hand, the image quality of the decoded secret image is acceptable when the secret image increases to 5, i.e., the PSNR is 33.01 dB. On the other hand, with the increasing number of images to hide, the decoded result shows a declining trend. The potential reason is the learning difficulty of the decoder increases with the increase of secret images as the maximum allowable magnitude of the universal perturbation is fixed. Appendix \ref{supp:secret_num} provides some visualization examples.

\textbf{Influence of the secret image types.} To investigate whether the proposed method can hide different types of images, we randomly choose two natural images from the ImageNet validation set and generate two images consisting of Chinese and English text. We rerun the proposed method on S2W and average the result decoded by different secret keys. Table \ref{tab:ablation_image_type} reports the evaluation results. As we can observe, the more complicated the secret image, the worse the decoded image quality is. The potential reason is that it is hard for the decoder to accurately learn the rich texture in natural images and intensive text in text images. More advanced model architecture and loss function may solve this problem. Nonetheless, we provide the visualization example in Appendix \ref{supp:secret_type}.
\section{Conclusion}
This paper proposes a novel universal perturbation-based secret key-controlled data hiding method. We treat the single universal perturbation as the secret data carrier to hide multiple secret images, which can be added to most cover images. Then, we extract different secret images by the devised decoder using different secret keys from one container image. Moreover, we devise a suppress loss function to map wrong secret keys to nonsense secret images, effectively protecting secret images from leakage. Extensive digital and physical experiments demonstrate the effectiveness of our method.
{
    \small
    \bibliographystyle{ieeenat_fullname}
    \bibliography{main}
}

\clearpage
\setcounter{page}{1}
\maketitlesupplementary

\setcounter{section}{0}
\setcounter{figure}{0}
\setcounter{table}{0}

\section{Robustness evaluation}
\label{supp:robust}
In this section, we provide detailed quantitative and qualitative results for the image quality of secret images decoded from the corrupted container images.

\subsection{Quantitatively result}
To perceive the decoding discrepancy of using the different secret keys, we provide the evaluation result of the secret image using different secret keys in Table \ref{tab:appendix_blur} and Table \ref{tab:appendix_jpeg}, respectively. As we can observe, on the one hand, in Table \ref{tab:appendix_blur}, Key-Stanford performs better on the S2W dataset than Key-UCLA, resulting in about a 3 dB lead in PSNR. In contrast, Key-UCLA performs better on the ImageNet dataset than Key-Stanford, resulting in about a 2 dB lead in PSNR. Such observation suggested that the imbalanced sampling of two secret keys makes the decoder prefer a specific secret key. Thus, a balanced sampling strategy should be introduced to solve this issue to avoid imbalanced decoding results. On the other hand, in Table \ref{tab:appendix_jpeg}, Key-Stanford performs better on both datasets consistently than Key-UCLA, resulting in about a 2 dB lead in PSNR. This observation suggests the robustness discrepancy of the decoder toward different corruptions, which indicates that it is hard to obtain better robustness against all corruptions. 

\begin{table*}[]
\setlength\tabcolsep{2pt}
\caption{Quantitatively results of secret images decoded from the corrupted of different kernel sizes(ks) of Gaussian blur.}
\label{tab:appendix_blur}
\begin{tabular}{c|c|c|ccccccccccccc}
\hline
\multirow{2}{*}{}        & \multirow{2}{*}{ks} & \multirow{2}{*}{Dataset} & \multicolumn{6}{c}{Key-Stanford}         &             & \multicolumn{5}{c}{Key-UCLA}              &          \\ \cline{4-9} \cline{11-16} 
                         &                     &                          & SSIM$\uparrow$            & PSNR$\uparrow$           & FID$\downarrow$            & LPIPS$\downarrow$              & APD$\downarrow$            & MSE$\downarrow$                &           & SSIM$\uparrow$            & PSNR$\uparrow$           & FID$\downarrow$            & LPIPS$\downarrow$              & APD$\downarrow$            & MSE$\downarrow$               \\ \hline
C-C$'$                   & \multirow{2}{*}{-}  & S2W                      & 0.8949 & 35.49 & 15.25  & 0.007365 & 3.51  & 0.000283 & & 0.8949 & 35.49 & 15.25 & 0.007365 & 3.5   & 0.000282 \\
                         &                     & ImageNet                 & 0.8955 & 35.48 & 13.1   & 0.002614 & 3.51  & 0.000283 & & 0.8955 & 35.48 & 13.1  & 0.002614 & 3.51  & 0.000283 \\ \cline{1-16} 
\multirow{10}{*}{M-M$'$} & \multirow{2}{*}{-}  & S2W                      & 0.9928 & 35.51 & 24.33  & 0.000937 & 2.42  & 0.000293 & & 0.9908 & 37.02 & 8.68  & 0.0011   & 2.08  & 0.000231 \\
                         &                     & ImageNet                 & 0.9954 & 38.53 & 11.91  & 0.000164 & 1.95  & 0.000179 & & 0.9941 & 39.43 & 5.55  & 0.000176 & 1.68  & 0.000164 \\ \cline{2-16}
                         & \multirow{2}{*}{3}  & S2W                      & 0.9578 & 28.96 & 51.18  & 0.004035 & 5.23  & 0.002694 & & 0.8946 & 26.48 & 50.86 & 0.007989 & 7.49  & 0.006305 \\
                         &                     & ImageNet                 & 0.9525 & 29.22 & 57.45  & 0.001423 & 5.44  & 0.002863 & & 0.9589 & 31.44 & 31.09 & 0.001056 & 3.98  & 0.00146  \\ \cline{2-16}
                         & \multirow{2}{*}{5}  & S2W                      & 0.9424 & 28.38 & 51.49  & 0.005021 & 6.3   & 0.004144 & & 0.84   & 25.01 & 85.25 & 0.011328 & 11.37 & 0.012707 \\
                         &                     & ImageNet                 & 0.8154 & 25.28 & 106.42 & 0.004346 & 13.61 & 0.016053 & & 0.8446 & 27.38 & 87.72 & 0.00373  & 10.82 & 0.010083 \\ \cline{2-16}
                         & \multirow{2}{*}{7}  & S2W                      & 0.954  & 28.99 & 50.85  & 0.004146 & 5.74  & 0.003349 & & 0.8618 & 25.53 & 71.44 & 0.00997  & 9.56  & 0.009453 \\
                         &                     & ImageNet                 & 0.8326 & 25.53 & 96.34  & 0.004088 & 12.68 & 0.014471 & & 0.858  & 27.71 & 78.13 & 0.003359 & 9.99  & 0.008721 \\ \cline{2-16}
                         & \multirow{2}{*}{9}  & S2W                      & 0.9588 & 29.26 & 50.1   & 0.003767 & 5.21  & 0.002669 & & 0.8686 & 25.67 & 67.46 & 0.009586 & 9.19  & 0.008845 \\
                         &                     & ImageNet                 & 0.8331 & 25.53 & 97.46  & 0.004126 & 12.64 & 0.014344 & & 0.8681 & 27.92 & 73.72 & 0.00316  & 9.3   & 0.007583 \\ \hline
\end{tabular}
\end{table*}

\begin{table*}[]
\setlength\tabcolsep{2pt}
\caption{Quantitatively results of secret images decoded from the corrupted of different quality(Q) factors of JPEG.}
\label{tab:appendix_jpeg}
\begin{tabular}{c|c|c|ccccccccccccc} 
\hline
\multirow{2}{*}{}        & \multirow{2}{*}{Q}    & \multirow{2}{*}{Dataset} & \multicolumn{6}{c}{Key-Stanford}           &          & \multicolumn{6}{c}{Key-UCLA}                        \\ \cline{4-9} \cline{11-16} 
                         &                       &                           & SSIM$\uparrow$            & PSNR$\uparrow$           & FID$\downarrow$            & LPIPS$\downarrow$              & APD$\downarrow$            & MSE$\downarrow$                &           & SSIM$\uparrow$            & PSNR$\uparrow$           & FID$\downarrow$            & LPIPS$\downarrow$              & APD$\downarrow$            & MSE$\downarrow$               \\ \hline
C-C$'$                   & \multirow{2}{*}{-}    & S2W                      & 0.8949 & 35.49 & 15.25 & 0.007365 & 3.5   & 0.000282 && 0.8949 & 35.49 & 15.25 & 0.007365 & 3.5  & 0.000282 \\
                         &                       & ImageNet                 & 0.8955 & 35.48 & 13.1  & 0.002614 & 3.51  & 0.000283 && 0.8955 & 35.48 & 13.1  & 0.002614 & 3.51 & 0.000283 \\ \cline{1-16} 
\multirow{12}{*}{M-M$'$} & \multirow{2}{*}{-} & S2W                      & 0.9928 & 35.51 & 24.33 & 0.000937 & 2.42  & 0.000293 && 0.9908 & 37.02 & 8.68  & 0.0011   & 2.08 & 0.000231 \\
                         &                       & ImageNet                 & 0.9954 & 38.53 & 11.91 & 0.000164 & 1.95  & 0.000179 && 0.9941 & 39.43 & 5.55  & 0.000176 & 1.68 & 0.000164 \\ \cline{2-16}
                         & \multirow{2}{*}{10}   & S2W                      & 0.838  & 18.87 & 82.46 & 0.012516 & 13.45 & 0.013484 && 0.8684 & 22.21 & 58.68 & 0.009048 & 8.52 & 0.006558 \\ 
                         &                       & ImageNet                 & 0.9583 & 29.43 & 28.01 & 0.001046 & 5.48  & 0.003244 && 0.9381 & 29.53 & 43.75 & 0.001603 & 5.67 & 0.003933 \\ \cline{2-16}
                         & \multirow{2}{*}{30}   & S2W                      & 0.9761 & 28.82 & 31.56 & 0.002077 & 4.33  & 0.001595 && 0.961  & 27.39 & 15.63 & 0.00264  & 4.02 & 0.001878 \\
                         &                       & ImageNet                 & 0.9848 & 33.75 & 14.2  & 0.000386 & 3.36  & 0.001063 && 0.9766 & 34.12 & 19.84 & 0.000627 & 3.18 & 0.001114 \\ \cline{2-16}
                         & \multirow{2}{*}{50}   & S2W                      & 0.9839 & 30.61 & 26.75 & 0.001453 & 3.67  & 0.001038 && 0.9705 & 28.67 & 13.05 & 0.002012 & 3.47 & 0.001395 \\
                         &                       & ImageNet                 & 0.9922 & 35.04 & 10.49 & 0.000207 & 2.75  & 0.000461 && 0.9867 & 35.39 & 13.48 & 0.000369 & 2.51 & 0.000443 \\ \cline{2-16}
                         & \multirow{2}{*}{70}   & S2W                      & 0.9854 & 31.18 & 25.2  & 0.001328 & 3.49  & 0.000894 && 0.9747 & 29.67 & 12.27 & 0.001715 & 3.2  & 0.001136 \\
                         &                       & ImageNet                 & 0.9937 & 35.46 & 9.75  & 0.000168 & 2.58  & 0.000351 && 0.9887 & 35.65 & 11.93 & 0.000319 & 2.38 & 0.000365 \\ \cline{2-16}
                         & \multirow{2}{*}{90}   & S2W                      & 0.9875 & 31.77 & 27.2  & 0.001236 & 3.37  & 0.000754 && 0.9715 & 29.79 & 14.64 & 0.002064 & 3.41 & 0.001145 \\
                         &                       & ImageNet                 & 0.9906 & 33.42 & 13.58 & 0.000258 & 3.05  & 0.000589 && 0.9757 & 31.94 & 19.72 & 0.000642 & 3.34 & 0.000886 \\ \hline
\end{tabular}
\end{table*}

\subsection{Qualitatively result}
\begin{figure*}[t]
	\centering
	\begin{minipage}{.8\linewidth}
		\centering
		\includegraphics[width =1.\linewidth]{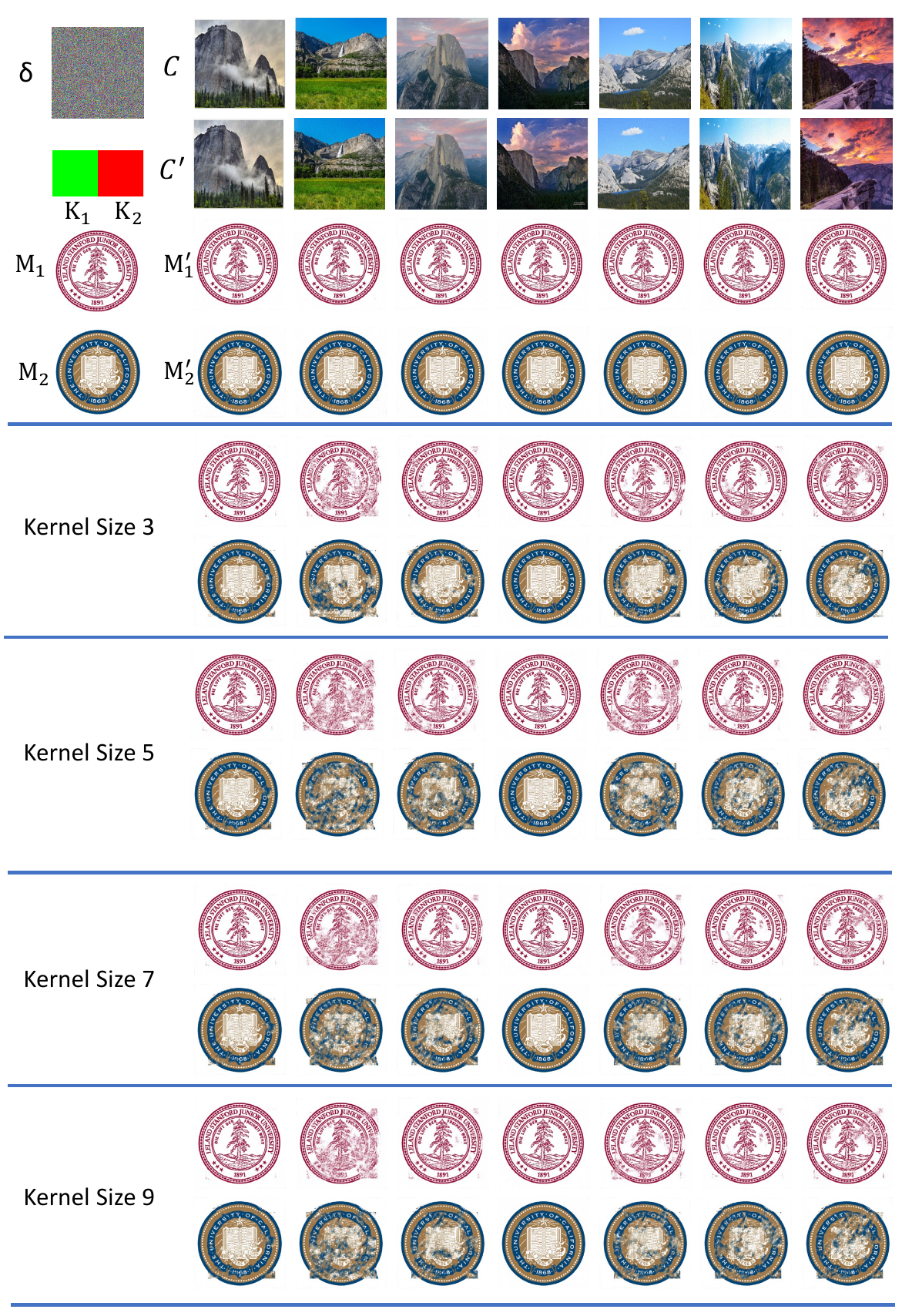}
	\end{minipage}
\caption{Visualization examples of secret images decoded from the corrupted of different kernel sizes of Gaussian blur. }
\label{fig:appendix_blur}
\end{figure*}

\begin{figure*}[t]
	\centering
	\begin{minipage}{.7\linewidth}
		\centering
		\includegraphics[width =1.\linewidth]{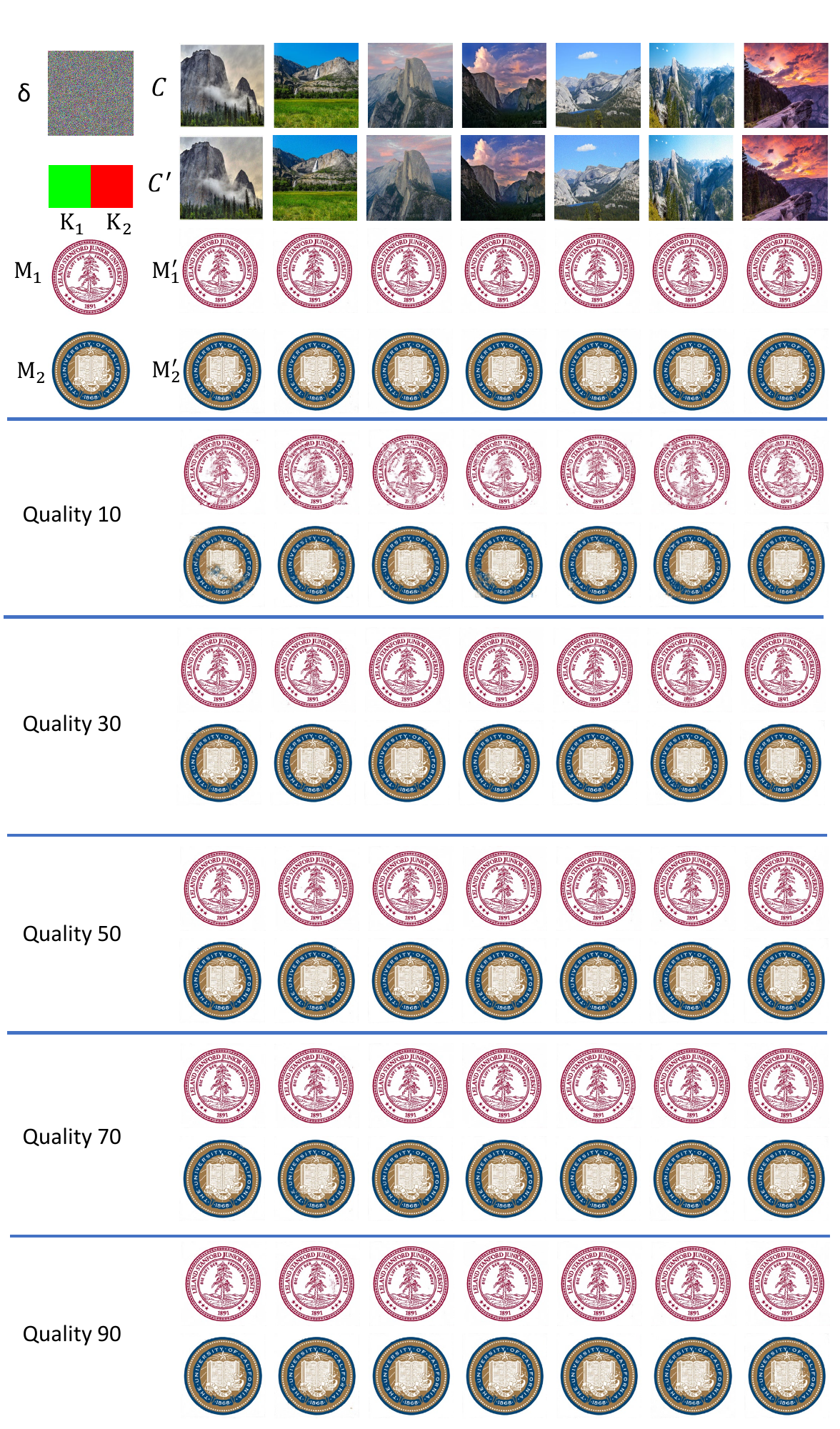}
	\end{minipage}
\caption{Visualization examples of secret images decoded from the corrupted of different quality factors of JPEG. }
\label{fig:appendix_jpeg}
\end{figure*}

To intuitively show the image quality of the decoding results, we provide the visualization examples of the decoded secrete image under different corruptions in Figure \ref{fig:appendix_blur} and Figure \ref{fig:appendix_jpeg}. As we can see, firstly, the visual quality of the secret images decoded from the JPEG compressed container image is better than the Gaussian blur one. Secondly, the visual quality of the secret images decoded from slightly corrupted container images is better than the severity one. In addition, although the image quality of most secret images decoded from the severity corrupted container image is worse, the content of the secret image is recognizable. Interestingly, we find that the image quality of some secret images decoded from the severity corrupted container image is fine, which is promising for the real application.

\section{Qualitatively result of the physical test}
\label{supp:physical_test}
\begin{figure*}[ht]
	\centering
	\begin{minipage}{1.\linewidth}
		\centering
		\includegraphics[width =1.\linewidth]{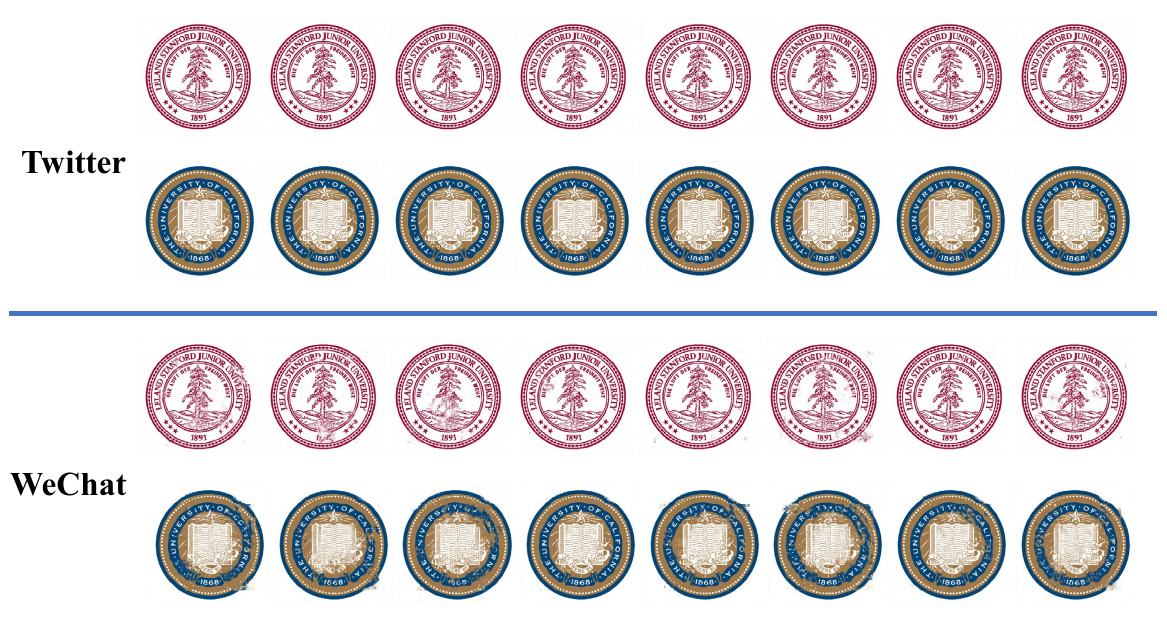}
	\end{minipage}
\caption{Visualization examples of secret images decoded from the image saved from Twitter and WeChat. }
\label{fig:appendix_physical}
\end{figure*}

In this section, we provide some visualization examples of the secret image decoded from social platforms in Figure \ref{fig:appendix_physical}, i.e., WeChat and Twitter. As we can see, the image quality of the decoded secret image is preserved well despite some detail is lost, and the image content is recognizable, which suggest that the proposed method shows potential usabilty in the real world.

\section{Effectiveness of the suppress loss function}
\label{supp:loss_supp}
\begin{figure*}[t]
	\centering
	\begin{minipage}{1.\linewidth}
		\centering
		\includegraphics[width =1.\linewidth]{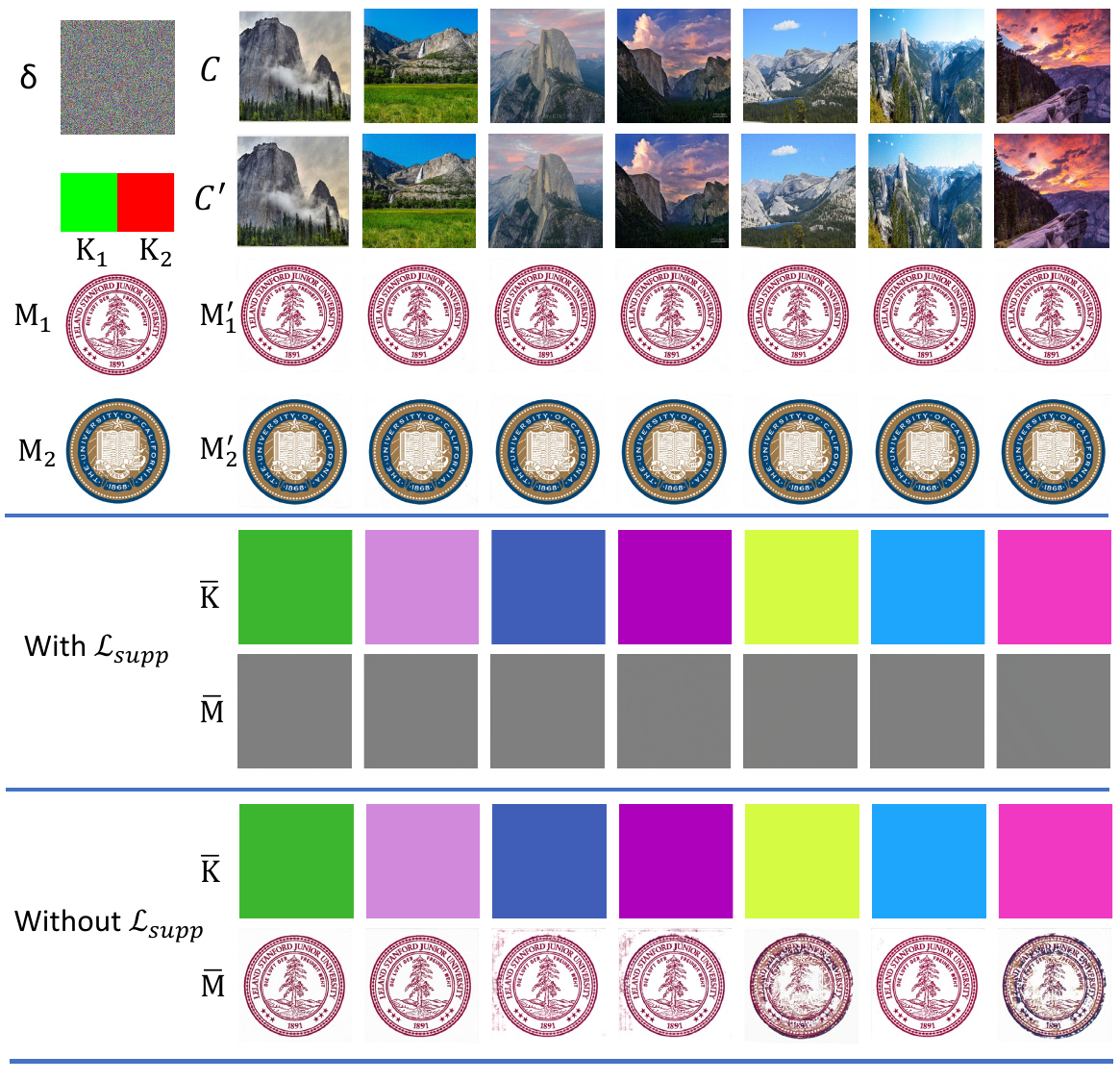}
	\end{minipage}
\caption{Visualization comparison sampled from the S2W dataset between with or without $\mathcal{L}_{supp}$. The notion in the figure indicates that universal perturbation $\delta$, cover image C, container image C$C'$, the secret key image K$_1$, and K$_2$, and the corresponding secret image M$_1$ and M$_2$, and the $\overline{M}$ and $\overline{M}$ indicates the illegal secret key and the corresponding decoded result. The figure above the first and second blue lines is the result with $\mathcal{L}_{supp}$. The figure under the second blue line is the result without $\mathcal{L}_{supp}$. Note that we only report the decoded result using the illegal secret key image for simplicity. }
\label{fig:loss_supp}
\end{figure*}

In this section, we provide intuitive visualization examples in Figure \ref{fig:loss_supp} with or without using the $\mathcal{L}_{supp}$  to make qualitative analysis. As show in Figure \ref{fig:loss_supp}, on the one side, the single universal perturbation is added to multiple cover images to construct the container image. Then different secret images with high quality can be extracted by the decoder using different legal keys, which demonstrates that our method can realize the secret-key controlled universal perturbation for data hiding. On the other hand, the devised loss function plays a significant role in preventing the secret image from leakage. Specifically, the decoder can not extract the accurate secret image using the random illegal secret key under the condition of with $\mathcal{L}_{supp}$. In contrast, without using $\mathcal{L}_{supp}$, although the image quality is not satisfying, the secret image is extracted by the illegal secret key. 

\section{Influence of the Secret Image Number}
\label{supp:secret_num}
\begin{figure*}[t]
	\centering
	\begin{minipage}{1.\linewidth}
		\centering
		\includegraphics[width =1.\linewidth]{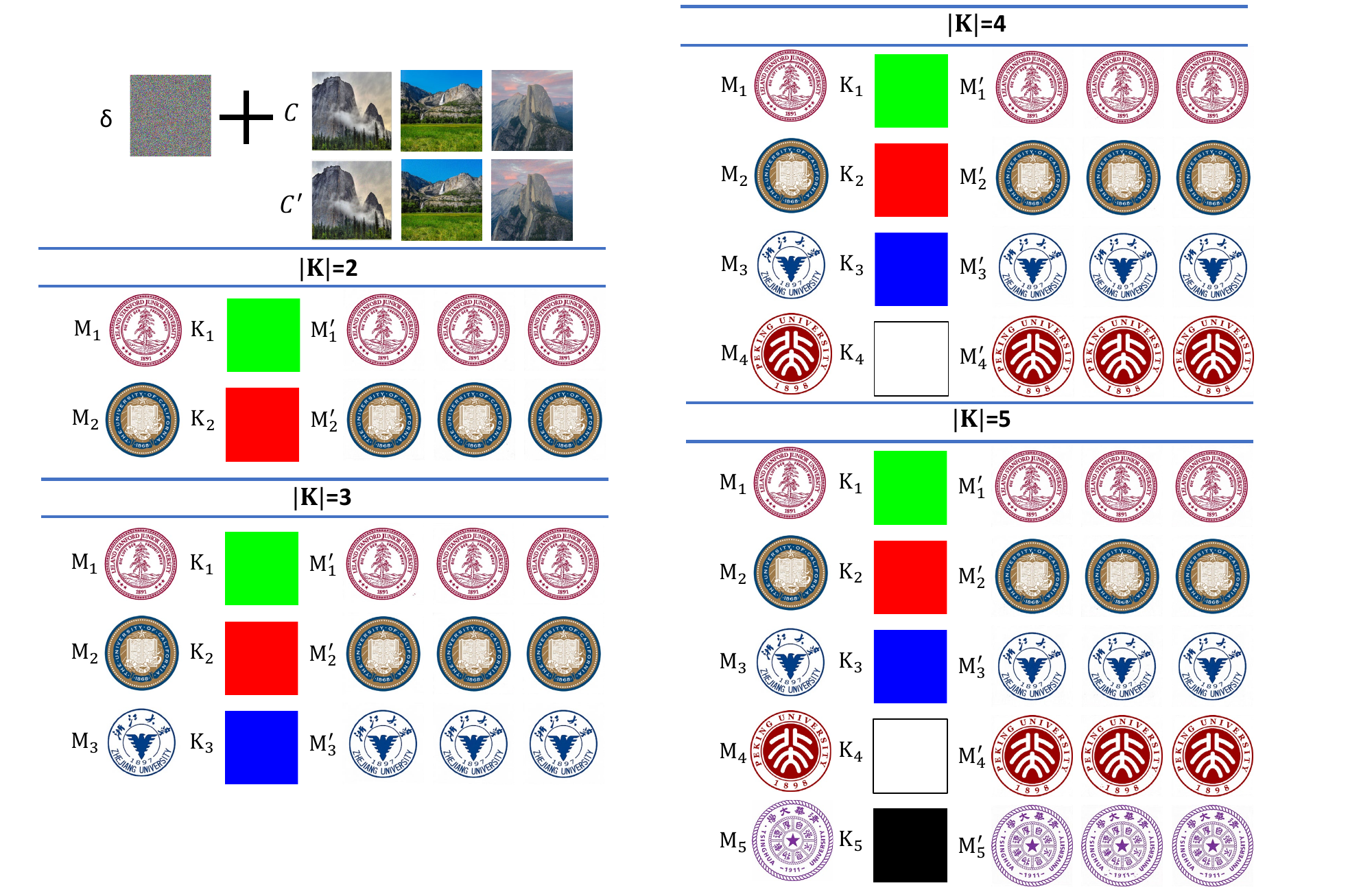}
	\end{minipage}
\caption{Visualization examples of different secret image number. $|K|$ denotes the number of secret images to be hidden in the universal perturbation. Zoom in for details.}
\label{fig:appendix_secret_num}
\end{figure*}

In this section, we investigate how many secret images can be hidden in a single universal perturbation with a fixed magnitude. As reported in the main text, the image quality of the secret image number decreases with the crease of the secret image number while the image quality is acceptable when the image number increases to 5. To provide more intuitive results, we provide some decoded secret images by using different secret image in Figure \ref{fig:appendix_secret_num}. As we can see, the image quality of the decoded secret image is preserved well with the increasing number of secret images to hide, while the allowable maximum magnitude of the universal perturbation is unchanged. Therefore, we speculate the potential reason is the considerable representational ability of the decoder. Nonetheless, the robustness of the decoder may weaken with the increase in the number of secret images, since the decoder needs to focus on decoding the secret image.

\section{Influence of Different Secret Image Types}
\label{supp:secret_type}
\begin{figure*}[t]
	\centering
	\begin{minipage}{1.\linewidth}
		\centering
		\includegraphics[width =1.\linewidth]{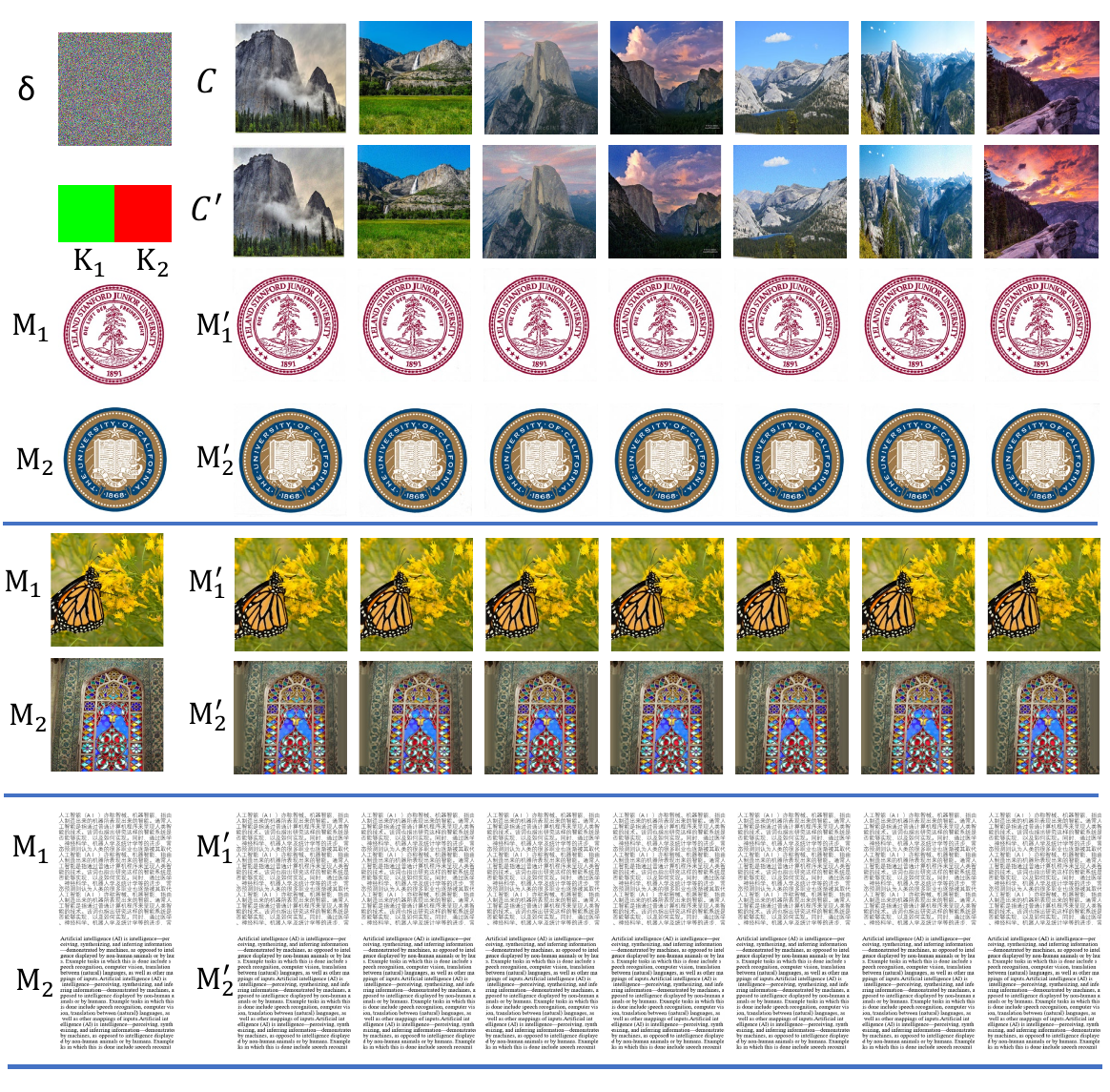}
	\end{minipage}
\caption{Visualization examples of different secret image types, i.e., school badge, nature image, and text image. Zoom in for details.}
\label{fig:appendix_secret_type}
\end{figure*}

The quantitative result in the manuscript shows that the proposed method is susceptible to the secret image type, where the rich texture (nature ) image limits the decoder's capability. However, the main content of the decoded secret image is still recognizable. We provide some visual examples in Figure \ref{fig:appendix_secret_type}. As we can observe, the proposed method performs well in badge and text images, where the decoded secret images are clearly identifiable. At the same time, the image quality of the decoded secret image decoded by using two secret keys shows a discrepancy. Specifically, the flower secret image exhibits better visual quality, while the rich texture image windows show worse, especially on both sides of the image, where there is a fine texture that the decoder hard to learn. 

\end{document}